Towards Global Earthquake Early Warning with the MyShake Smartphone Seismic Network

Part 2 - Understanding MyShake performance around the world

Authors: Qingkai Kong, Robert Martin-Short, Richard M. Allen

Corresponding author: Qingkai Kong

Address: 209 McCone Hall, UC Berkeley, Berkeley, CA, 94720

Email: kongqk@berkeley.edu

**Abstract:**

The MyShake project aims to build a global smartphone seismic network to facilitate large-scale earthquake early warning and other applications by leveraging the power of crowdsourcing. The MyShake mobile application first detects earthquake shaking on a single phone. The earthquake is then confirmed on the MyShake servers using a "network detection" algorithm that is activated by multiple single-phone detections. In part two of this two paper series, we report the first order performance of MyShake's Earthquake Early Warning (EEW) capability in various selected locations around the world. Due to the present sparseness of the MyShake network in most parts of the world, we use our simulation platform to understand and evaluate the system's performance in various tectonic settings. We assume that 0.1% of the population has the MyShake mobile application installed on their smartphone, and use historical earthquakes from the last 20 years to simulate triggering scenarios with different network configurations in various regions. Then, we run the detection algorithm with these simulated triggers to understand the performance of the system. The system performs best in regions featuring high population densities and onshore, upper crustal earthquakes M<7.0. In these cases, alerts can be generated ~4-6 sec after the origin time, magnitude errors are within ~0.5 magnitude units, and epicenters are typically within 10 km of true locations. When the events are offshore or in sparsely populated regions, the alerts are slower and the uncertainties in magnitude and location increase. Furthermore, even with 0.01% of the population as the MyShake users, in regions of high population density, the system still performs well for earthquakes larger than M5.5. For details of the simulation platform and the network detection algorithm, please see part one of this two paper series.

**Introduction**

MyShake is an effort to turn user's smartphones into portable seismometers, which can be used to monitor and record earthquakes globally. The mobile application monitors the accelerometer inside the device and uses and artificial neural network (ANN) trained to detect earthquake motions. Following the release of the MyShake to the public in 2016, users from all over the world now contribute earthquake data, which can enable various seismological and civil engineering applications (see part one of this two paper series for references). One specific goal for the MyShake project is to build a global earthquake early warning system to reduce the earthquake hazards. Working towards this goal, in part one, we introduced the design of a new network detection algorithm and described a simulation platform that can be used to estimate and evaluate the performance of MyShake networks in different configurations. In the present paper, we conduct simulations for all historical earthquakes M > 4.0 since January $1^{st}$ 1980 for a range of earthquake-prone regions around the world including California, New Zealand, Nepal, Central America, Haiti and Sulawesi (Indonesia) in addition to several others shown in the supplementary material. The six regions we present in this paper were chosen because they represent a wide range of tectonic environments, population distributions and levels of socioeconomic development. Furthermore, each region has been affected by a major earthquake in recent years.

For each region, we assume that 0.1% of the total population have the MyShake application installed on their smartphones. This number is the approximate proportion of the population of Los Angeles (LA) region that have download the MyShake application to date. These simulations allow us to assess the general performance of the network detection workflow in each region with this density of users, determine the spatial and magnitude distribution of events that could be detected and report their expected warning times and location errors. Furthermore, we conduct a suite of 200 simulations of the single most damaging event in each region. This provides a distribution of errors in origin time, epicenter location and magnitude that might be expected for these events given different samplings of the population. We also describe the results of tests conducted in each region using MyShake user level of 0.01% of the population.

Overall, we show that at a penetration level of 0.1% of the population, MyShake would provide accessible, useful earthquake early warning to communities worldwide. At a penetration level of 0.01%, the network is still capable of accurately locating earthquakes, but will only successfully detect those that occur within close proximity to dense urban areas.

**Overview of the regional simulations**

In order to assess the potential benefits of using MyShake networks for EEW on a global scale, we perform a suite of simulations using historical events around the world as shown in Figure 1. We highlight six regions of interest to discuss, results from all the other regions are listed in the supplementary materials. For each region, we use the MyShake trigger generation workflow to produce triggers for all events M > 4.0 from January 1$^{st}$ 1980 to April 1$^{st}$ 2019 assuming that 0.1% of the population is running the MyShake mobile app. The simulated triggers for each event are then provided to the network detection algorithm, which is instructed to report an estimate of the event parameters for the first alert. Using the historical catalog in this way allows us to assess the typical magnitude range and proximity to population centers of events in each region and thus comment on the performance of the system. Although a full exploration of the parameter space is required to determine the effects of specific features of a region have upon algorithm performance, our hope is that this analysis will provide a first order understanding of the MyShake performance in various regions.

Our six selected regions represent a wide range of tectonic settings, population distributions and levels of socioeconomic development. These three factors combine in a complex fashion to determine the usefulness of MyShake as an EEW system in each region. Each of the regions we choose has also experienced a major, damaging event in recent years. In order to assess the combined impact of the various uncertainties built into the population sampling, trigger generation workflows and the stability of the system performance, we conduct a suite of 200 simulations for each of these large events and report the distribution of location, magnitude and timing errors. This is important because the uncertainties we incorporate into our calculations of trigger times and ground accelerations at a device level represent our best understanding of how MyShake devices perform in practice. It is important to note that for simplicity we model these events as point sources, which ostensibly they are not, meaning that in practice the true distribution of ground acceleration values that would have been observed by a MyShake network may be quite different from our simple models. Nevertheless, the error distributions returned by the network detection algorithm provide us with some indication of the best and worst case performance scenarios for a major earthquake in each region.

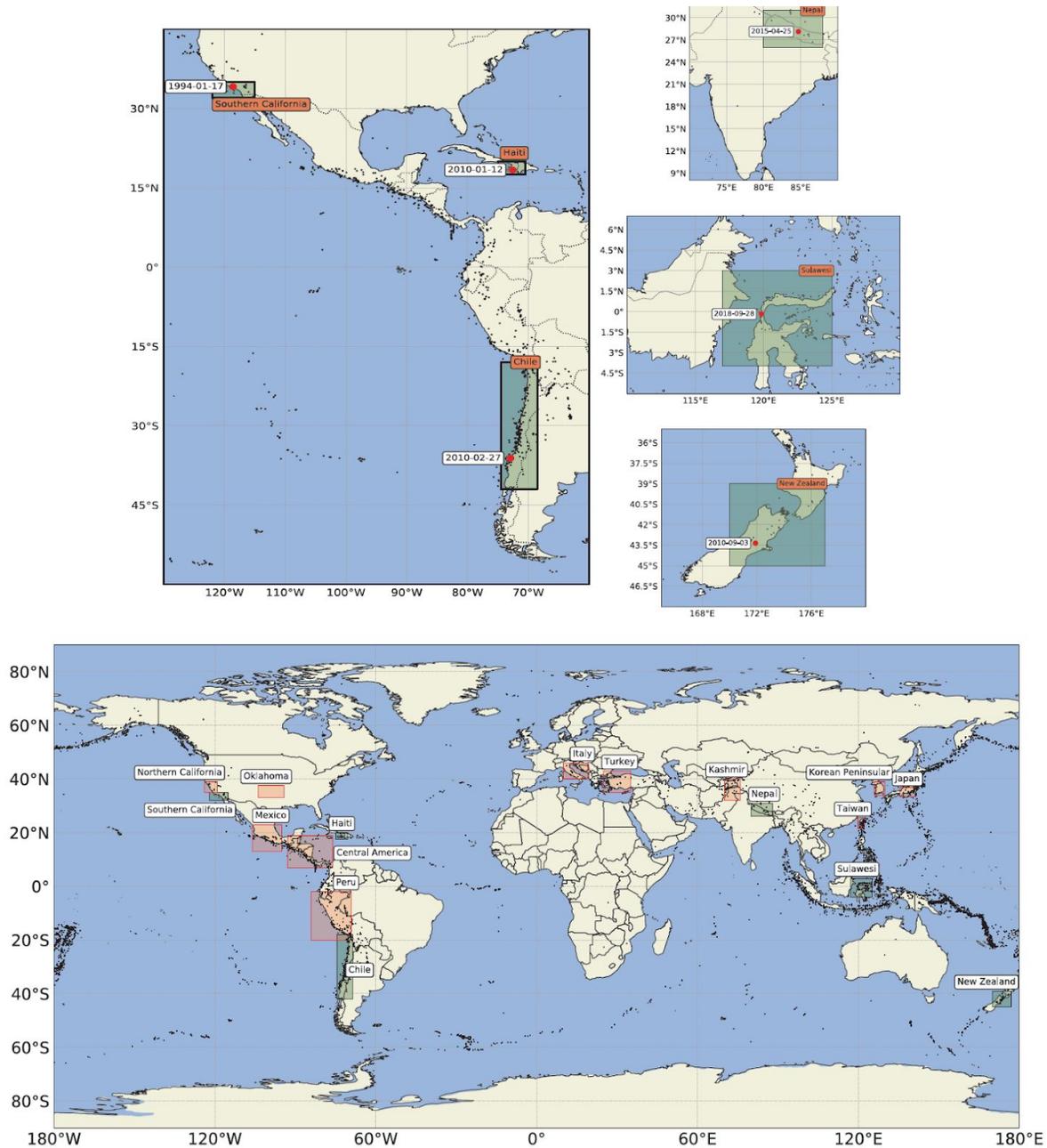

**Figure 1**. Regions of the world where we have tested using the MyShake simulation platform. The bottom panel shows all of the tested regions. Labelled boxes show the geographic extent of each region. The top panels provide an enlarged view of the six selected regions, which are chosen because they encompass major population centers and the sites of significant historical events, whose epicenters and dates are labelled. Together these regions also represent a wide range of tectonic regimes, population densities and socioeconomic development levels. Please see the supplementary materials for the results from regions boxed in red on the global map.

**Performance of the simulated historical events**

*Southern California*

Southern California is one of the world's most well-studied regions in relation to earthquake hazard. The presence of the on-land, strike-slip boundary between the Pacific and North American plate, which manifests itself in the San Andreas fault zone, is responsible for seismic activity here (Working Group on California Earthquake Probabilities, 1995). The seismogenic zone typically lies within the uppermost 15 km, and the region experiences a damaging event M>6.0 approximately every 10-20 years. The risk is especially high in the densely populated Los Angeles basin, through which the San Andreas Fault runs. Southern California is well instrumented, containing dense networks of traditional seismometers facilitating an operational EEW system called ShakeAlert that issues public alerts [(Given et al. 2014)](#).

Figure 2 shows MyShake simulation results for Southern California. The network performs relatively well here: most events have epicentral distance errors of less than 15 km, and those with a larger number of triggers appear to be located with greater accuracy, as expected. The observed scatter in the magnitude-error plot (Figure 2b) can be attributed to the difficulty of estimating magnitudes from single ground acceleration values due to the fact that earthquakes of a range of sizes produce broad and overlapping distributions of ground motion (e.g. Boore *et al.*, 2014). Nevertheless, most events have magnitude errors of less than one unit. The time to first alert varies between 2.3 and 14.8 seconds, reflecting the spatial distribution of events and population centers shown in Figure 2a. As indicated by the histogram in 9e, a MyShake network generated by 0.1% of the population would be sufficient to detect almost all events M>5.0 in this region. Many of the smaller events go undetected because they do not generate a sufficient number of triggers.

The histograms in Figure 3 show error distributions from 200 simulations of the 1994 M6.7 Northridge event. See also movies S4 and S5 for examples of the network detection algorithm's performance with this event. While the majority of errors in Figure 3 are small, there is a tail of large magnitude, distance and time errors that correspond to instances where the event is initially very poorly located due to variability in the locations of the randomly selected MyShake phones, and uncertainties incorporated into the generated trigger data. Subsequent updates performed as the simulation proceeds act to remove these extreme error values, as illustrated in Figure S1.

When the sampled population is dropped to 0.01% in this densely populated area, the results are not dramatically different. A smaller proportion of earthquakes are detected, although the majority of those M>5.5 are still captured. An obvious difference is the increase of the time to the first alert, especially for the events that are far from population and have fewer phones nearby (the red dots in the magnitude vs time to first alert panel). The maximum time from origin of the earthquake to the first alert increased from ~14s for the 0.1% case to ~20s. See figures in the supplementary material for more information.

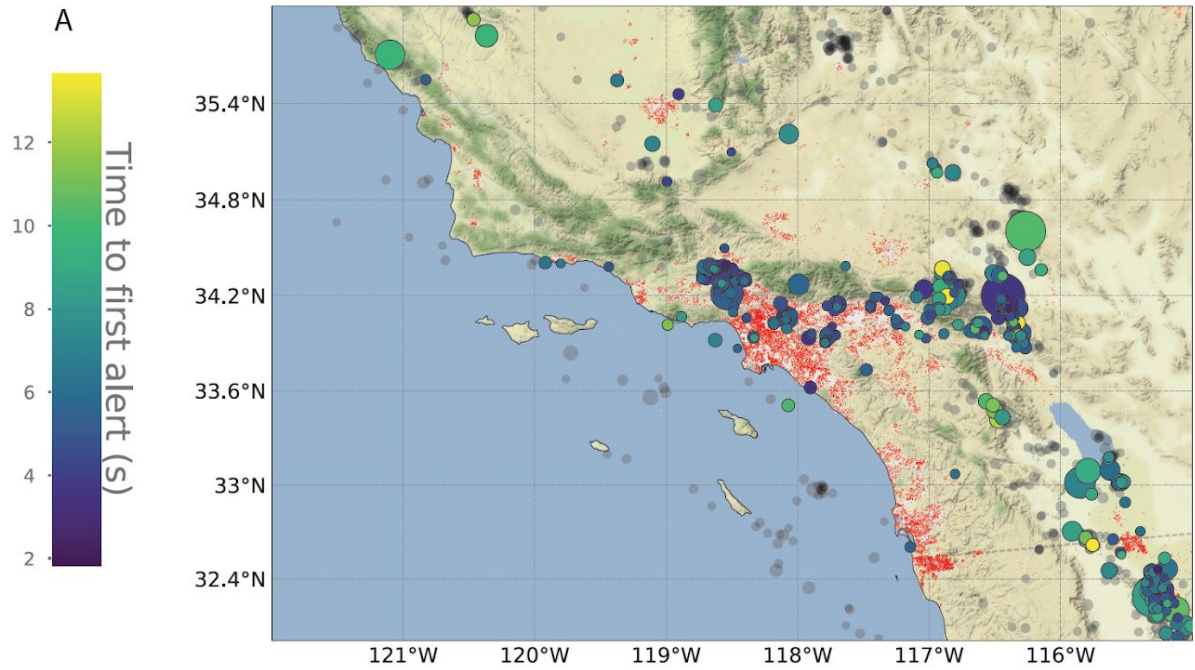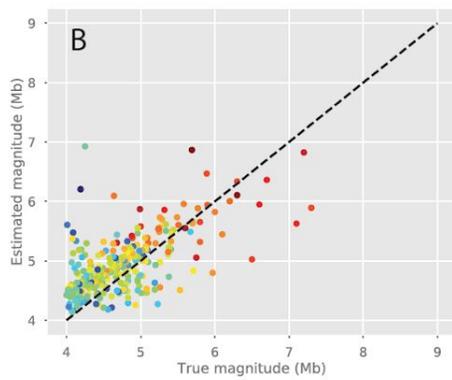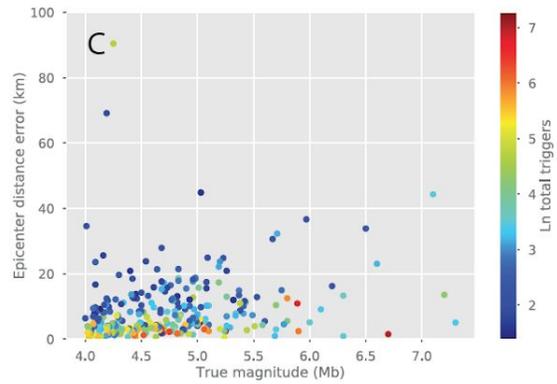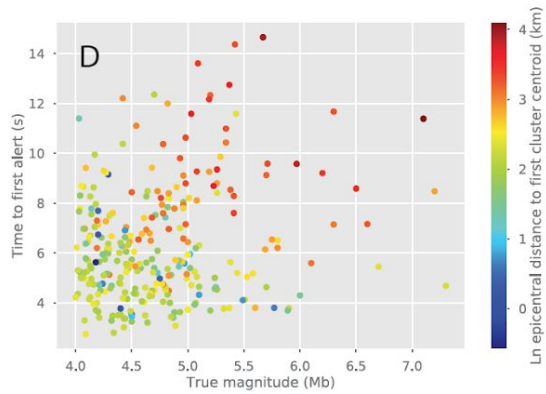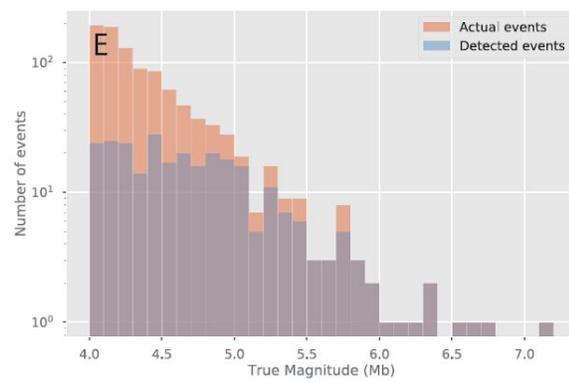

**Figure 2.** Summary of simulation results for Southern California. (a) Map of the study region showing all events M>4.0 since 1980. Epicenters are colored by the time to first alert from origin time of the earthquake. Red dots show the location of simulated MyShake users, representing 0.1% of the total population. Gray circles are earthquakes that were not detected by the MyShake network: They are typically small, remote or offshore. (b) MyShake estimated magnitude versus USGS catalog magnitude. Earthquakes are colored by the fraction of total triggers occuring on the P-wave in the simulation. Typically larger events are associated with a greater fraction of P-wave triggers. (c) Plot of location error against magnitude. Events are colored by the natural log of the total number of triggers illustrating that the larger location errors are for events with fewer triggers (because they are far from population centers). (d) Time to first alert as a function of magnitude. Points are colored to indicate the natural log of the epicentral distance to the centroid of the first trigger cluster as identified by DBSCAN. This serves as an indication of the distance from the epicenter to the closest population center. We observe that smaller distances are associated with faster alerts. (e) shows the proportion of catalog events that are successfully detected by the MyShake network as a function of magnitude.

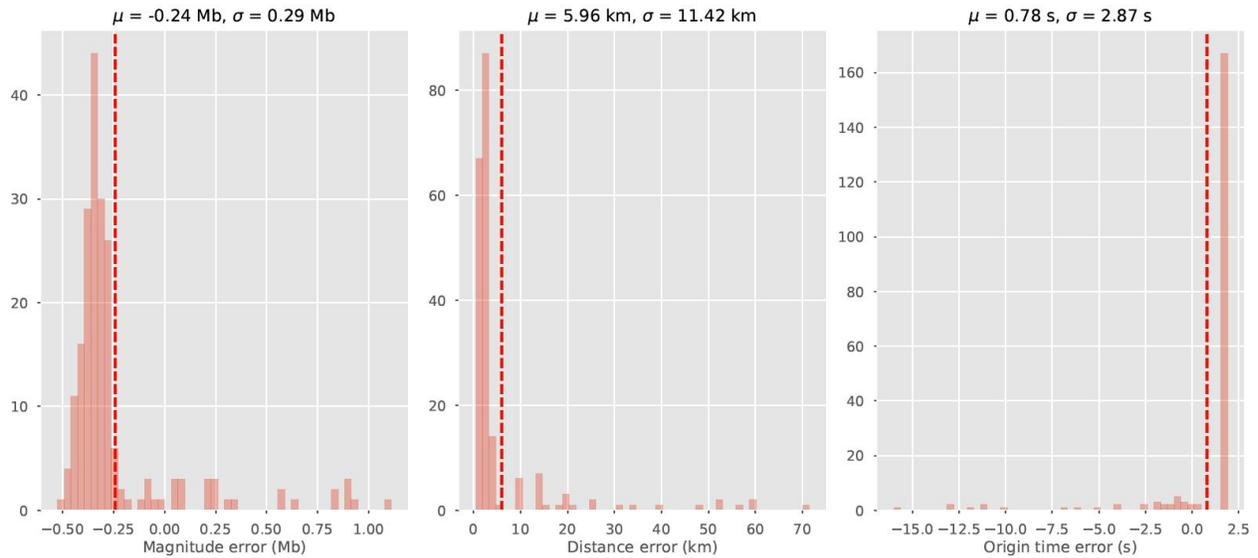

**Figure 3.** Results of 200 simulations of the January 17, 1994, M6.7 Northridge earthquake showing the distribution of magnitude, location and origin time errors. These errors are calculated using the estimated values minus the catalog values.

*Haiti*

The earthquake hazard in Haiti was brought to international attention by the devastating M7.0 event of 2010, which killed over 200,000 people and caused major economic damage (Bilham, 2010). In addition to being one of the most densely populated Caribbean nations, Haiti is the poorest country in the western hemisphere and suffered from weak to non-existent building regulation prior to 2010 (Bilham, 2010; Frankel *et al.*, 2011).

Haiti makes up roughly half of the island of Hispanola, which lies within a zone of deformation caused by motion of the Caribbean and North American plates (DesRoches *et al.*, 2011). Haiti is the site of two major strike-slip faults: the Enriquillo Fault, on which the 2010 event occurred, which crosses densely populated regions in the south of the country and the Septentritional fault, which lies along the northern coast. Both have produced several major historical earthquakes (DesRoches *et al.*, 2011). Haiti is also threatened by earthquakes occurring in two subduction zones: The Puerto-Rico Trough to the northeast and the Muertos Trough to the southeast. A traditional seismic network consisting of seven instruments was set up in the country following the 2010 earthquake, but is arguably too sparse to be useful for EEW. However, despite widespread poverty, mobile phone use in Haiti is ubiquitous and growing rapidly.

MyShake simulation results for Haiti are shown in Figure 4. The MyShake network performs well in this case in terms of earthquake detectability, magnitude and location accuracy. This is due to the high population density, shallow earthquakes and onshore faults. All but one of the M>4.0 events have epicentral distance errors of less than 20km and the scatter in magnitude estimates is small. Furthermore, the vast majority of all events M>4.5 are successfully detected, including all onshore M>4.0 events in Haiti and most in neighboring Puerto Rico. Most of the alerts are issued within 4-8 sec of the event origin time.

In the specific case of the 2010 event, it is located with distance errors of less than 6 km in 198 of the 200 test simulation runs, with a mean time to first detection of 4.7 seconds and mean magnitude error of 0.4 units (Figure 5). These results suggest that with 0.1% of the population downloading MyShake, the system could provide an effective and much needed EEW tool for the people of Haiti. More broadly, this example highlights the potential benefits that MyShake could bring to developing earthquake-prone nations that have ubiquitous smartphone use.

With a sample size of 0.01% of the population (see the corresponding figure in the supplementary material), a smaller proportion of events are detected compared with the 0.1% counterpart, but for those that are the results are still very encouraging.

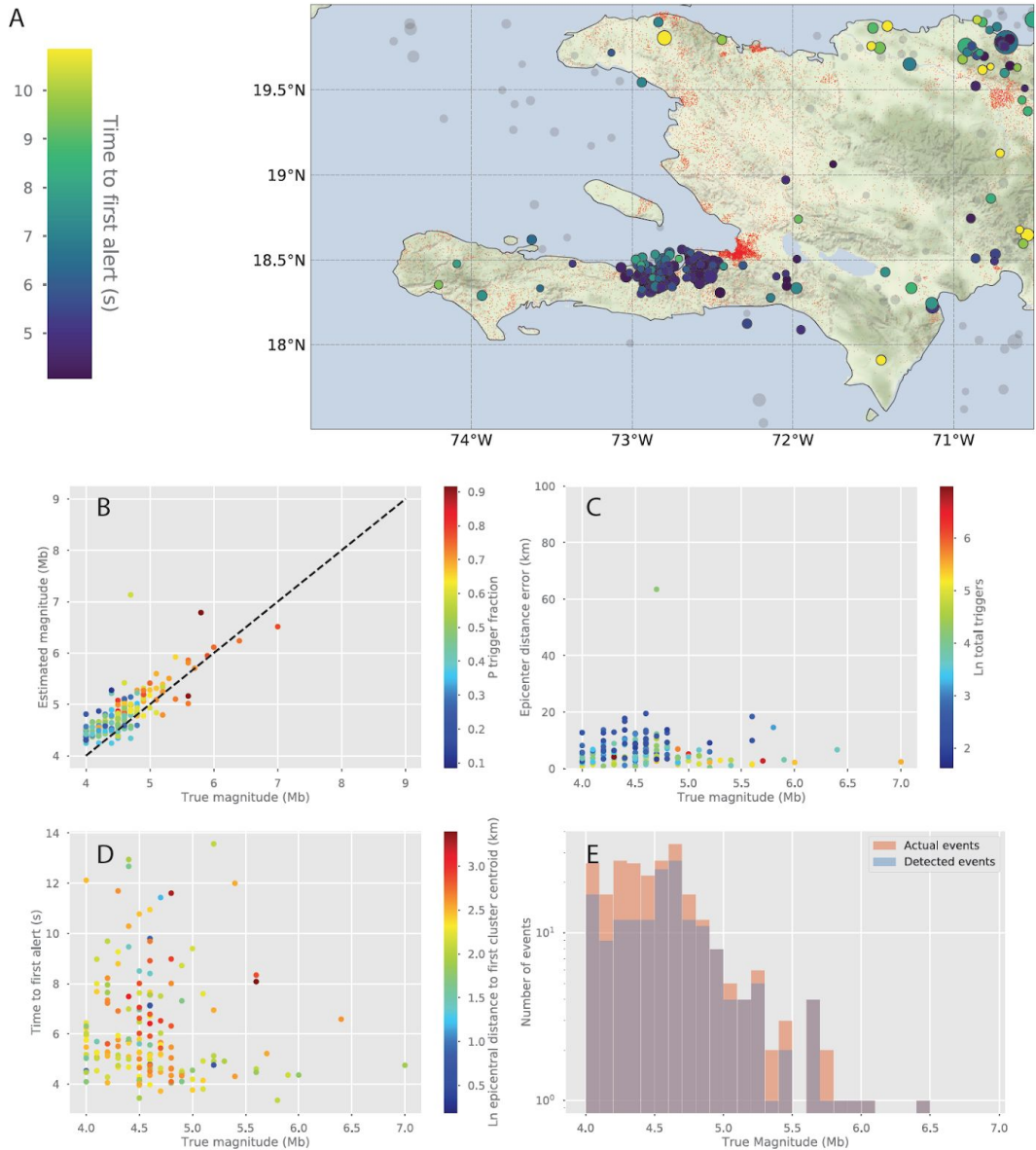

**Figure 4.** Summary of the simulation results for Haiti, with the panels arranged in the same format as in Figure 2. The network detection algorithm performs very well in this region,

detecting the majority of M>4.0 events and locating them accurately in time, space and magnitude. As seen in (a), many of the detected events are aftershocks of the 2010 M7.0 earthquake, and occur in very close proximity to densely populated regions of the country.

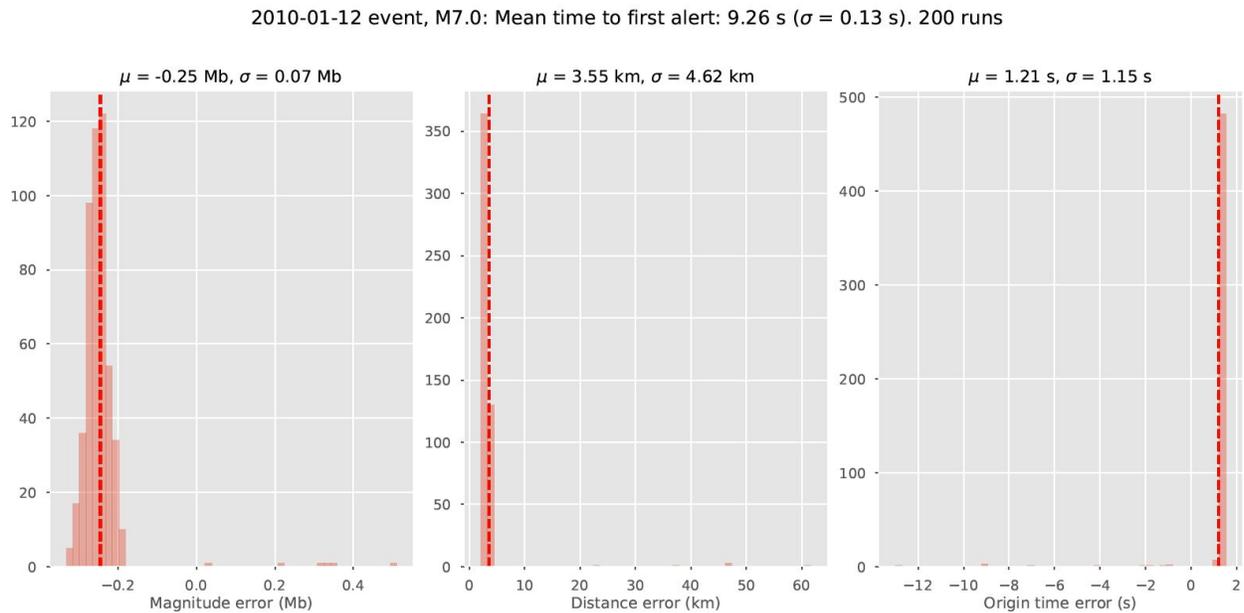

**Figure 5.** Results of 200 simulations of the January 12, 2010, M7.0 Haiti earthquake showing the distribution of magnitude, location and origin time errors.

*Chile*

Chile is one of the world's most seismically active regions. Tectonically, it is dominated by rapid suduction of the Nazca plate eastward beneath the South American plate, which varies from about 80mm/yr in the south of the country to 65mm/yr in the north. There are two main categories of earthquakes here; those occurring in the Wadati-Benioff zone, the largest of which tend to occur offshore near the Peru-Chile trench and those associated with deformation of the South American plate onshore in the forearc. Large subduction zone events here are frequent;

numerous M>8.0 events are known to have occurred within the past century, many of which were responsible for major loss of life and economic damage. Notable earthquakes include the May 22nd, 1960 M9.5 near Valdivia, the largest recorded earthquake in history, the February 27th, 2010 M8.8 in central Chile and the September 16th, 2015 M8.3 near Illapel, all of which also generated large tsunamis.

The Chilean population is accustomed to experiencing earthquakes and much has been learned from the county's long history of devastating events. This has produced a high level of public awareness of the threat, strict building codes for life-safety and a significant monitoring effort on the part of the country's National Seismological Center.

The population distribution varies dramatically along the length of Chile. About 85% of the population live in urban areas, mainly those associated with the cities of Santiago and Valparaiso in the central part of the country. Many of these urban areas are coastal, putting them at risk of strong shaking and tsunamis generated by megathrust earthquakes. Our chosen region for earthquake simulation spans the country from 42 deg S to 18 deg S, encompassing the vast majority of its population and sites of major historical earthquakes.

Figure 6 shows our results. They are generally poorer than in regions dominated by crustal onshore fault hazard such as Southern California and Haiti. Figure 6a shows that most events of M > 4.0 in close proximity to urban areas are detected, with events occurring further offshore generally having larger times to first alert. Figures 6(b-d) show significant scatter in the magnitude and location errors, especially for smaller events. This pattern is also seen in simulations for other subduction zones (namely Central America, Mexico and Japan) and is due to the challenging event-population geometry in these settings, (i.e. out-of-network events) and also that many of these events are deeper than 10 km, which the depth is fixed in the simulations. Furthermore, the Cua & Heaton (2009) magnitude relations were not intended for use in subduction settings. This point is emphasised by the error histograms shown in Figure 7, which were created by running 200 simulations of the February 27th 2010, M8.8 event. The large magnitude errors clearly indicate that the Cua & Heaton (2009) relationship is saturated for events of this size. Despite these concerns, we suggest that MyShake networks could still be a valuable asset for issuing rapid warnings to urban areas in Chile, even if the magnitude is underestimated. When the sampled population is dropped to 0.01%, the network detection algorithm is only able to detect shallow crustal events of M~4.5-7.0 occurring in close proximity to urban areas. The results are encouraging for these earthquakes, but larger, offshore events are typically not detected because they are too distant from regions with sufficiently dense clusters of devices.

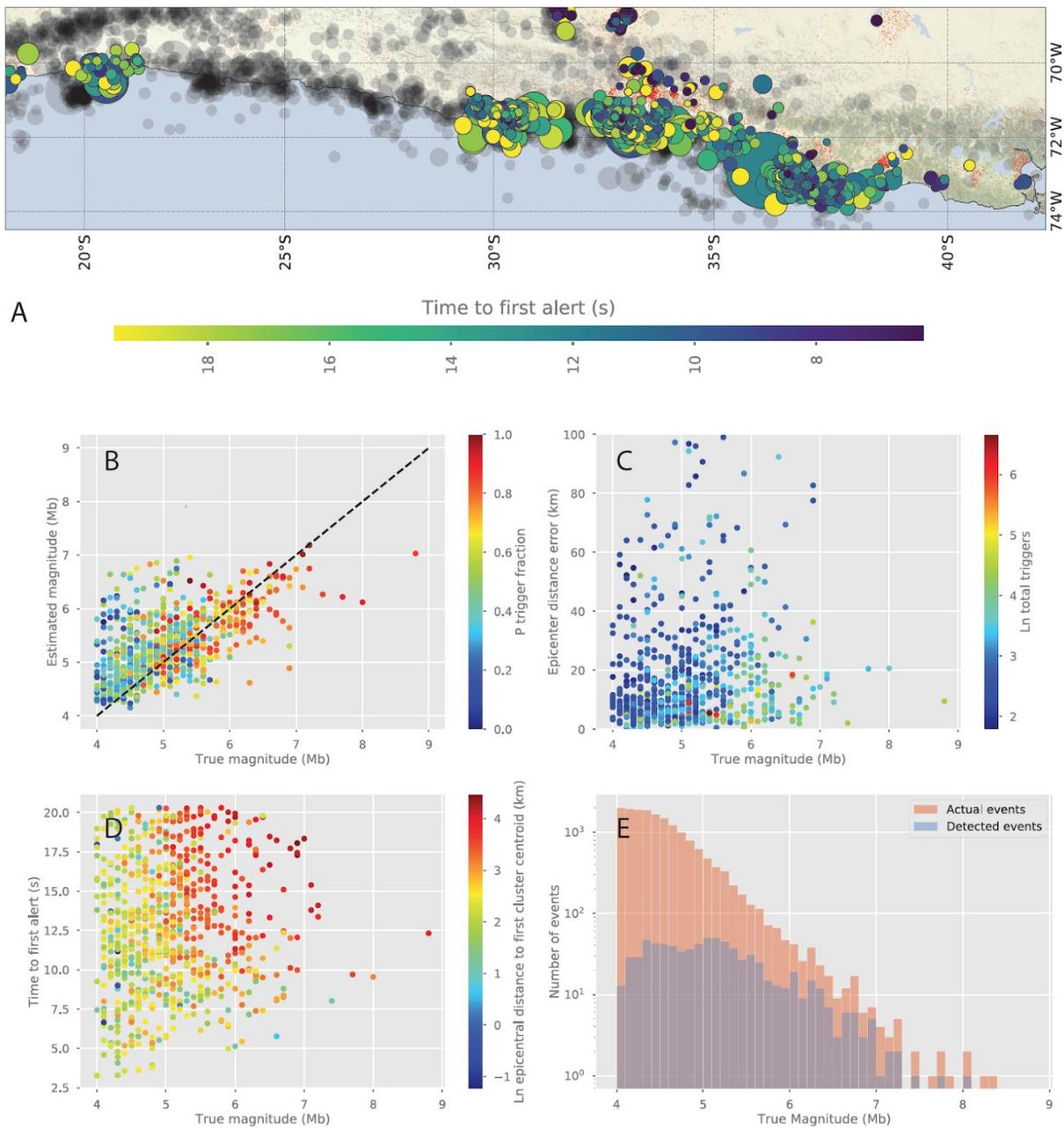

**Figure 6.** Summary of the simulation results for Chile. Chile is a classic subduction setting where the majority of the earthquakes occur offshore, to the west of population centers. This poses special challenges to the network detection algorithm, as reflected in its reduced performance here.

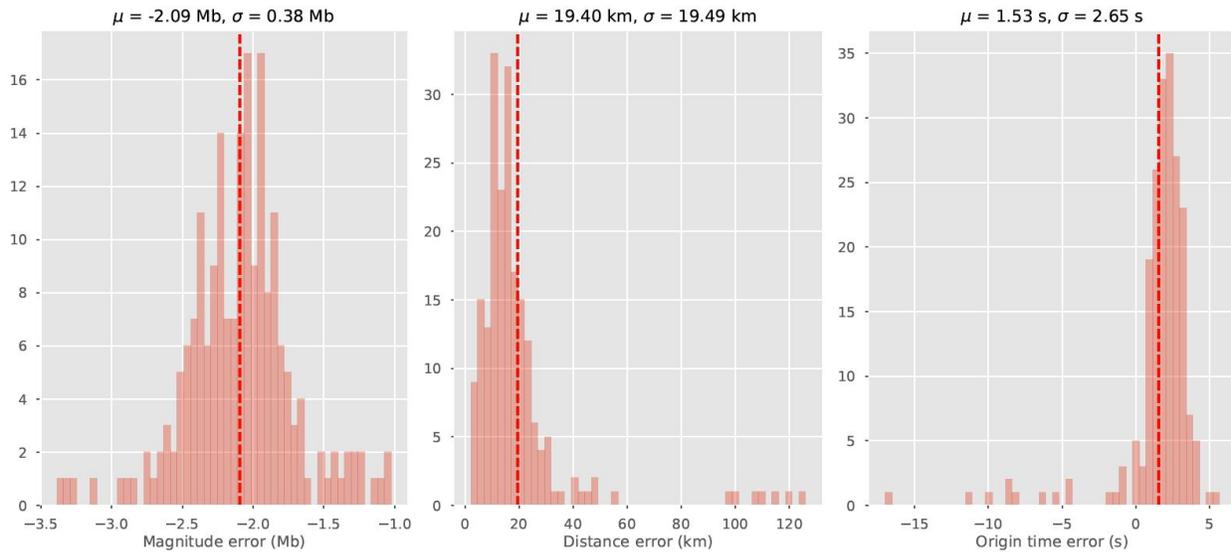

**Figure 7.** Results of 200 simulations of the February 27th 2010 M8.8 central Chile earthquake, which is the largest earthquake we attempted to simulate in this study. Our magnitude scaling relationships and point source approximation are poorly suited to such an event.

*Nepal*

Seismic hazard is also high in Nepal and northern India, which contain some of the world's most densely populated urban centers. These are at risk from major earthquakes occurring on thrust faults along the southern margin of the Himalayan mountains (Bilham *et al.*, 2001). Most of the strain accumulated by India's 20mm/yr convergence with southern Tibet is thought to be released by large earthquakes, with notable examples having occurred in 1803, 1833, 1897, 1895, 1932, 1950 and 2015. The April 25th 2015 M7.8 event near Kathmandu killed almost 9000 people and brought international attention to the region. As in Haiti, many of the casualties occurred due to the collapse of poorly constructed buildings in densely populated urban areas, most notably in Kathmandu. The National Seismological Center of Nepal operates a small network of traditional sensors in the country, which could potentially be used for EEW. However, mobile phone use in Nepal and northern India is already ubiquitous, with smartphone penetration at currently more than 50% (NepaliTelecom, 2019).

Figure 8 summaries our simulations in Nepal. Most of the detectable earthquakes occur in a narrow band just south of the Himalayas; those occurring north of the mountains in Tibet are typically too far from population centers to be detected. Our results here are somewhat similar to those for Southern California; most detected events have epicentral distance errors of less than 20km. In contrast to Southern California, the magnitudes of events in Nepal appear to be consistently overestimated by about 0.5 Mb. Our 200 simulations of the 2015 M7.8 earthquake indicate that it is consistently mislocated by about 10 km and its magnitude is underestimated due to the saturation of magnitude from the relationship we are using (Figure 9). Nevertheless, given a mean time to first alert of ~7 seconds, this would have provided Kathmandu with ~16 seconds of warning. Furthermore, this performance could likely be improved by tuning the parameters of the network detection algorithm. As in the aforementioned regions, the alerting system remains useful when using just 0.01% of the population, but the proportion of actual events that are detected drops dramatically (see corresponding figures in the supplementary materials).

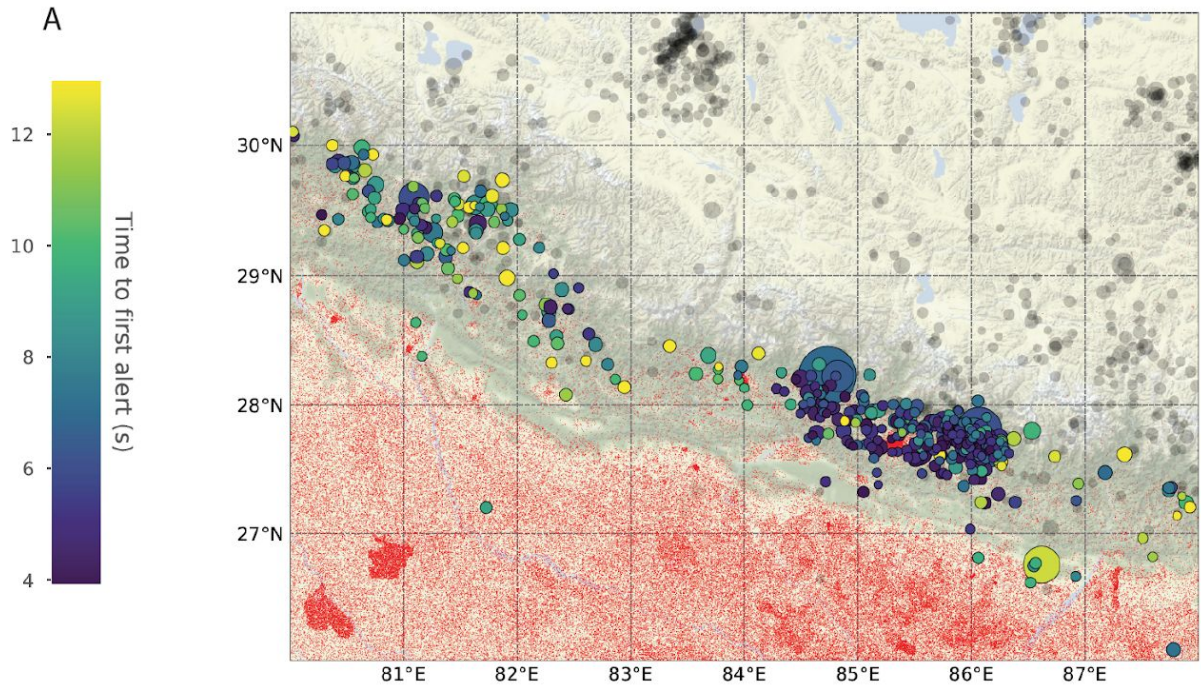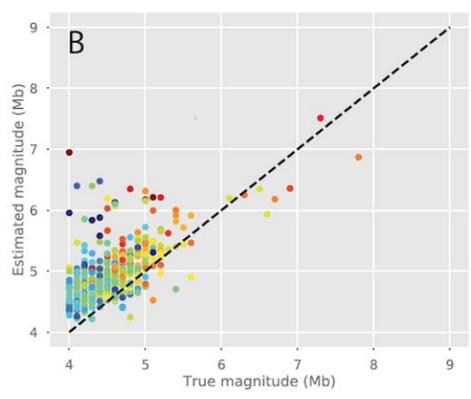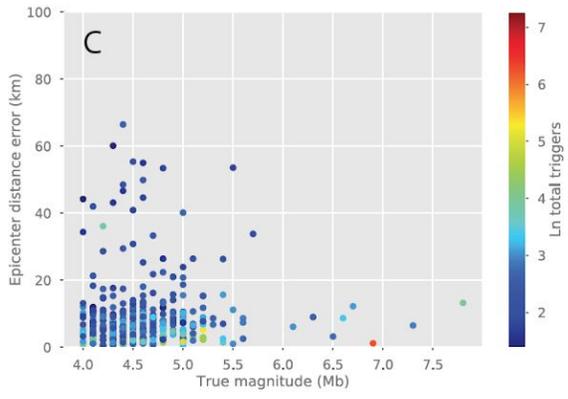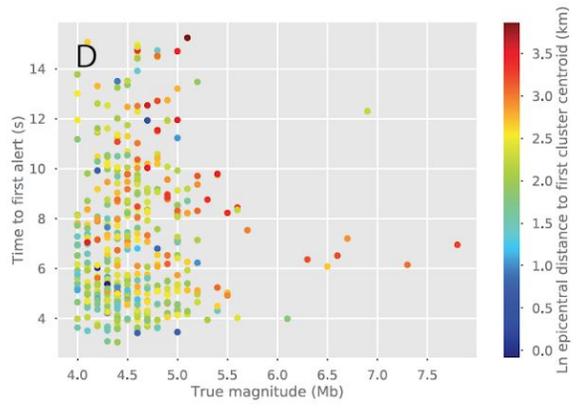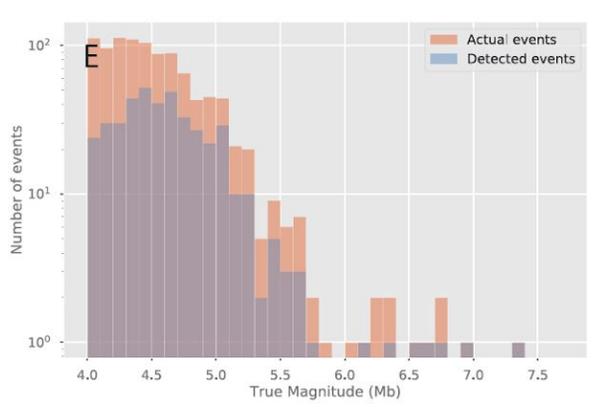

**Figure 8.** Summary of the simulation results for Nepal. Most M>4.5 events south of the Himalayas are detected, but those to the north are too distant from population centers to cause enough triggers.

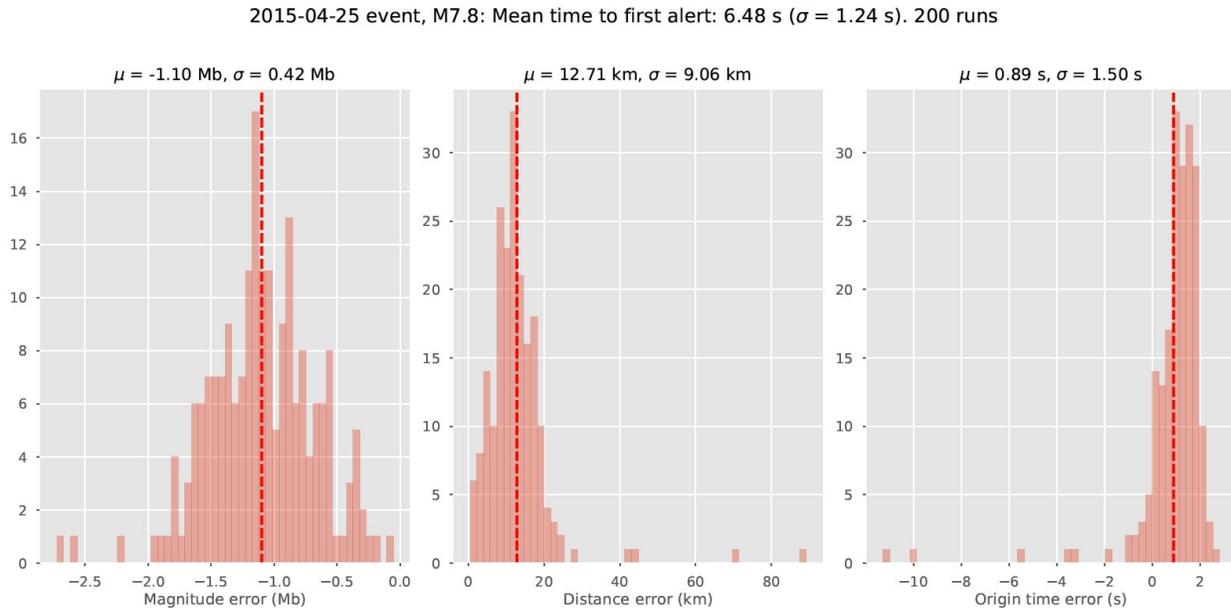

**Figure 9.** Results of 200 simulations of the April 25th 2015 M7.8 Nepal earthquake. The distributions of errors from these simulations are wider than in the case of Southern California and Haiti, suggesting that the population distribution relative to the event poses more of a challenge to the network detection algorithm here.

*Sulawesi*

Indonesia is one of the most seismically active countries in the world. Many of the islands formed through volcanism and accretion of terranes along the region's two major tectonic boundaries. These consist of northwards subduction of the Indo-Australian plate beneath the Eurasian plate in the south and west and a very complex boundary involving subduction of both the Pacific and Philippine Sea plates in the northwest (e.g. Villeneuve *et al.*, 2002 ). The island of Sulawesi lies just southwest of Borneo, near the center of the Indonesian archipelago and in a complicated and poorly understood tectonic setting that involves both subduction and strike-slip motion along the boundaries of several microplates (Villeneuve *et al.*, 2002). Sulawesi was chosen for MyShake simulations because it is relatively densely populated, exhibits a diversity of seismic activity and suffered devastation in the September 28th 2018 M7.5 Palu event. Despite having a predominantly strike-slip mechanism, this event generated a 2m high tsunami that inundated the coastal city of Palu and claimed over 4000 lives (Carvajal *et al.*, 2019). In addition to the 2018 event, Sulawesi has experienced three further tsunamigenic earthquakes during the past century and a host of damaging strike-slip events along the Palu-Koro and Matano faults, which cut though the island and lie close to urban areas (Carvajal *et al.*, 2019). Similar to Haiti and Nepal, Sulawesi has no operating earthquake early warning system but smartphone use is widespread and growing rapidly.

Our results from the Sulawesi simulations are shown in Figure 10. The colors in Figure 10b indicate that there are generally more triggers per event than in the case of Nepal, suggesting populations in closer proximity to the events. This observation is also supported by Figure 10c. The variations in the magnitude and location estimation performance is likely due in part to variations in the tectonic regime for different events. Events off the north coast of Sulawesi are typically in a subduction setting, while those running through the central part of the island are occuring along the Palu-Koro and Matano transform faults. The locations and alert times of this second group of events are especially promising. The simulation also performs well in the case of the September 28th 2018 M7.5 Palu event (Figure 11). Its large size inevitably means that its magnitude is underestimated due to the scaling relationship that we are using, but it is well located and an initial alert time of 6 seconds after the origin time would have provided about 15 seconds of warning before the S-wave reached the densely populated Palu region. When the number of MyShake phones used in the simulation is reduced to 0.01%, only events within close proximity of the urban areas of Palu and Gorontalo are detected (see corresponding figures in the supplementary materials). However, for those that are detected, their location, magnitude and timing errors are similar to what they are at larger sample sizes.

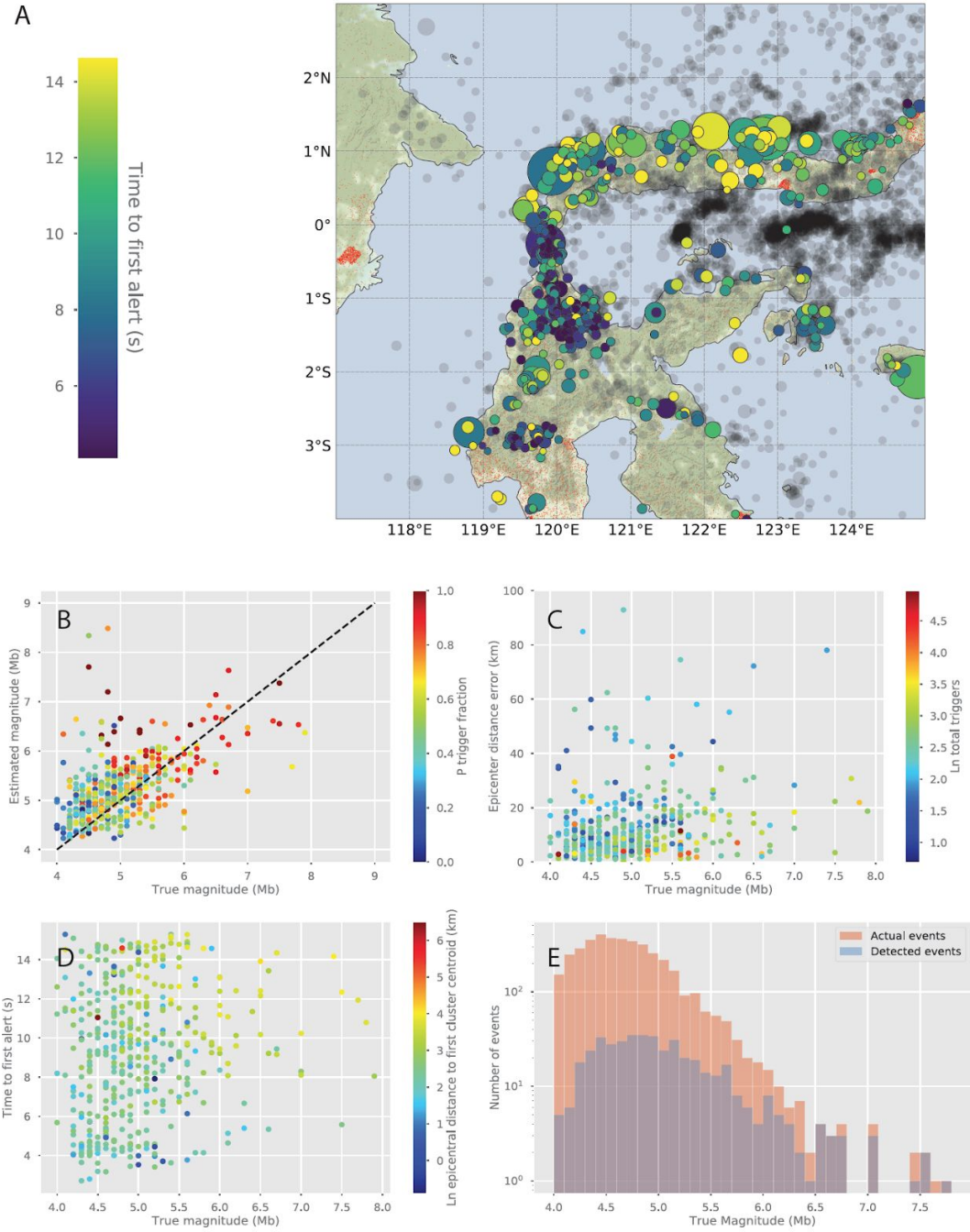

**Figure 10.** Summary of the results from the Indonesian island of Sulawesi. The performance is best for onshore earthquakes that occur near the densely populated Palu region, but events occurring just offshore along the northern coast are also reliably detected.

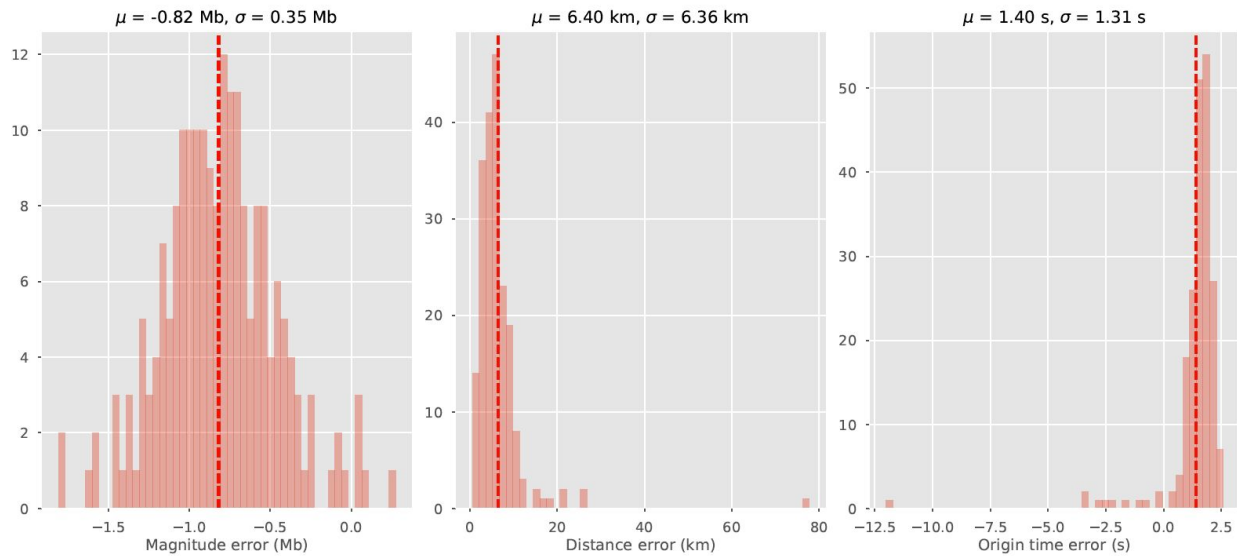

**Figure 11.** Results of 200 simulations of the September 28th, 2018, M7.5 Haiti earthquake showing the distribution of magnitude, location and origin time errors.

*New Zealand*

New Zealand is our final choice of demonstration regions for the MyShake simulation platform. Of the regions considered here, it is most similar to Southern California in terms of socioeconomic development, but has a much lower population density and thus poses unique challenges. In contrast to the aforementioned areas, earthquake fatalities here have historically been low, in part due to stringent construction regulations and a low population. Nevertheless, New Zealand is very seismically active. It sits astride a plate boundary that transitions from eastwards-verging subduction along the Hikurangi margin off the east coast of the North Island to left-lateral strike-slip motion along the Alpine Fault zone that bisects the South Island (Anderson and Webb, 1994). Four of the country's major cities, Christchurch, Wellington, Hastings and Napier, lie in close proximity to these major fault zones and have each suffered damage from M>7.0 events over the past century. In the aftermath of the 2011 M6.2

Christchurch event, which killed 185 people, there has been renewed interest in seismic monitoring and EEW in New Zealand (Wood *et al.*, 2012).

Figure 12 summarizes the simulation results for New Zealand. The low population density means that only a small proportion of the total number of events are detected by MyShake. However, the events that are detected are those that lie closest to population centers, most have epicentral distance errors of less than 20 km and magnitude errors of less than 0.5 magnitude units. The magnitude of some of the smaller events are overestimated; the majority of these small events are aftershocks of the February 22nd 2011 M6.1 Christchurch earthquake. In our 200 simulations of this event, the magnitude error distribution is approximately centered on zero and the location is generally accurate (Figure 13). The mean first alert time of 4 seconds after the origin time would not have been sufficient to issue warnings to central Christchurch, which only 10 km from the epicenter. Nevertheless, the city's suburbs and nearby town of Ashburton could have received several seconds of warning had this network been present. If the sampled population is dropped from 0.1% to 0.01%, only M>7 earthquakes and those occurring directly beneath Christchurch are detected. Again, however, their associated errors are not dramatically different to what they are when a higher sample population is used.

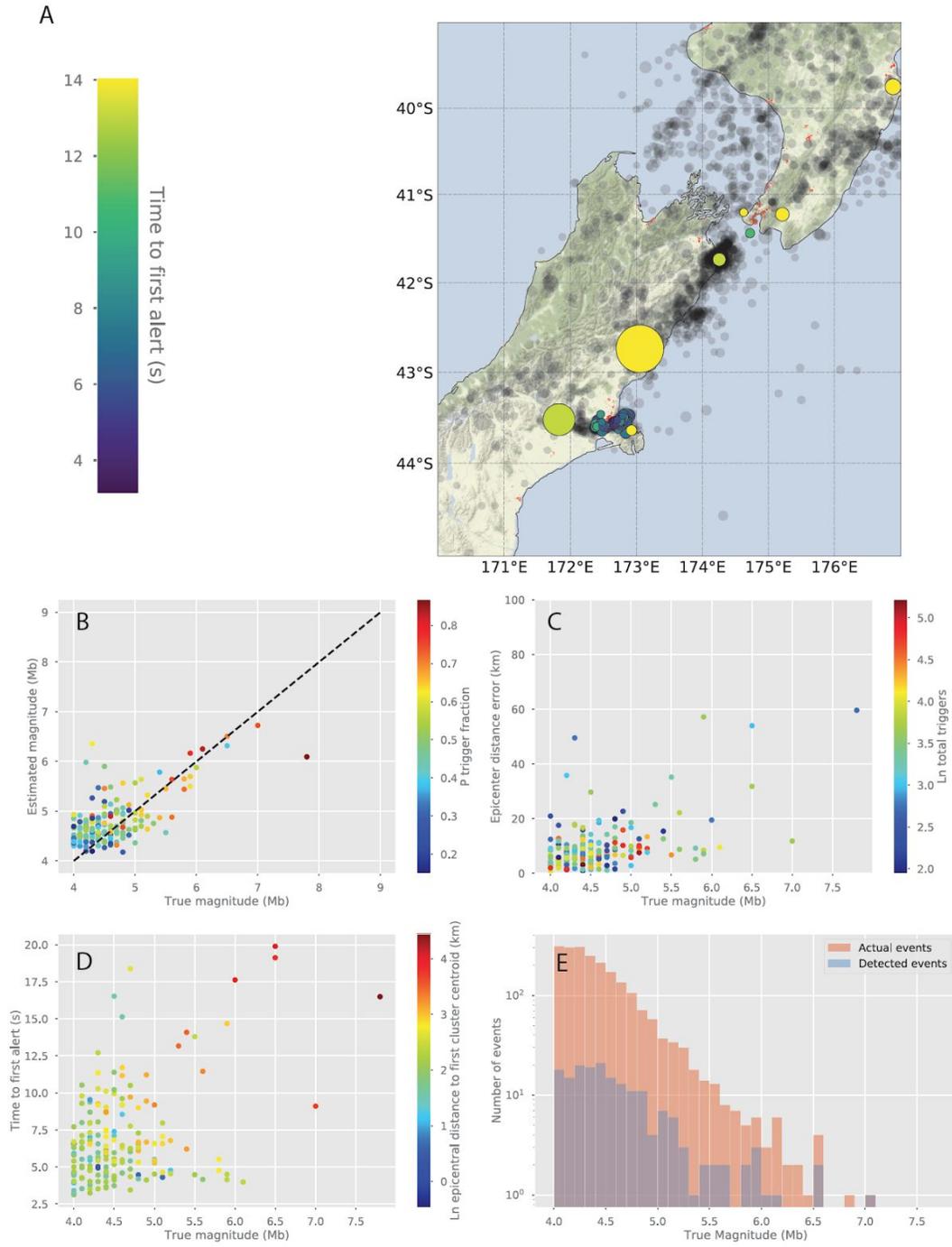

**Figure 12.** Summary of our simulation results for New Zealand. The region is sparsely populated, but events occurring close to urban areas are generally located accurately. Many of the M<4.5 events that are successfully detected are aftershocks of the 2011 Christchurch earthquake.

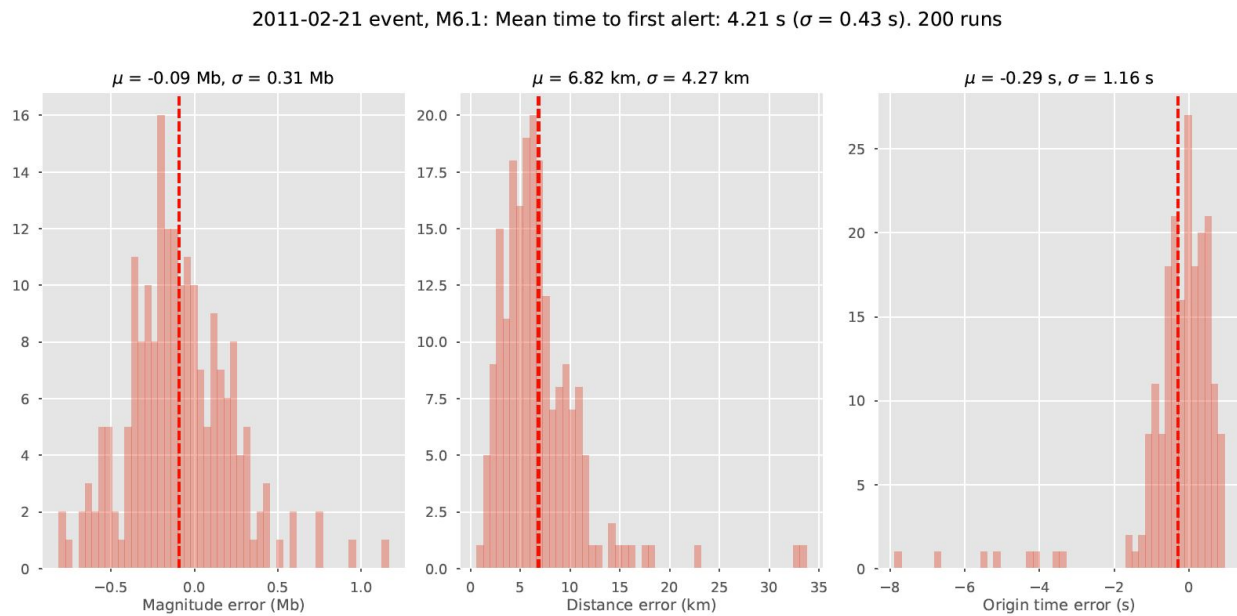

**Figure 13.** Results of 200 simulations of the February 22th, 2010, M6.1 Christchurch earthquake showing the distribution of magnitude, location and origin time errors.

**Discussion**

The results presented above provide a first order indication of the potential performance of a smartphone-based earthquake early warning system in various regional settings encompassing different tectonic environments and populations distributions. Our tests suggest that MyShake EEW performance is best in densely populated regions experiencing shallow, onshore crustal earthquakes in close proximity to urban areas. When these regions have poor building standards the potential benefits of an EEW system are even greater, as exemplified by the results for Haiti. Rapid MyShake event detection is more severely tested in subduction zone settings such as Chile, where the most destructive earthquakes tend to be very large and occur offshore. Evidently use of the Cua & Heaton (2009) amplitude-distance relationships to estimate the magnitude, and the approximation of the events as point sources breaks down in the case of such earthquakes, leading to poorer magnitude estimation. Nevertheless, despite this limitation the algorithm is still able to quickly and accurately locate these large earthquakes, even in the most challenging settings. Furthermore, we have built the simulation platform and network detection workflow such that it is straightforward to adjust parameters and test different methods of magnitude estimation. Thus, the algorithm could be tailored to each region in which it is deployed.

The fact that magnitude estimation performance varies considerably from region to region suggests that a more regionally tailored approach is necessary in future refinements of the algorithm. Additionally, we find that location and magnitude accuracy improves with increasing confidence in our ability to discern whether a trigger has occurred on a P or an S phase. Increasing the accuracy of the P vs. S classifier could also be a fruitful avenue for future development.

Our results also show that in general, events with more initial triggers tend to be located more accurately, as would be expected. As the simulations proceed beyond the initial detection stage and incorporate more and more updates, the total number of triggers increases further. Consequently event locations do generally improve and outlier error values, such as those seen in Figures 7, 9 and 11 for many of the regions, do disappear (Figure S1).

**Conclusion**

This two-paper series introduces a new detection algorithm and simulation platform for MyShake designed to understand its potential for earthquake early warning (EEW) on a global scale. In this part two of the series, we report the results of simulations in different parts of the world assuming 0.1% of the population are MyShake users. Due to the dynamic nature of this smartphone seismic network, the configuration of the network will be different at different times of the day or locations (Kong et al., 2019). Therefore, we simulated events in various regions using the past events occurring at different times of the day to gain insights of the performance of the system.

Detailed statistics of the performance in different regions are presented in this paper and in the supplement materials. We find that performance is best for onshore crustal earthquakes in regions with dense populations such as Sulawesi (Indonesia), Haiti and southern California. In these cases, alerts can be generated ~4-6 sec after the origin time, magnitude estimates are within ~0.5 magnitude units, and locations are within 10 km of true locations. When events are offshore such as in subductions zones, the alerts are slower and the uncertainties in magnitude and location increase. Also, in sparsely populated regions such as New Zealand, small and moderate magnitude events away from population centers may not be detected at all.

The assumption that 0.1% of the population downloads the MyShake app perhaps presents the greatest challenge for MyShake to be successful as a global earthquake early warning system. Our hope is that with an improved mobile application that offers a variety of features including EEW [(Rochford et al. 2018)](#), the network will grow sufficiently.

**Data and Resources**

The USGS Comcat catalog can be accessed at: https://earthquake.usgs.gov/fdsnws/event/1/. The data for the Gridded Population of the World can be accessed at

https://beta.sedac.ciesin.columbia.edu/data/set/gpw-v4-population-count-adjusted-to-2015-unwpp-country-totals. MyShake data are currently archived at Berkeley Seismology Laboratory and use is constrained by the privacy policy of MyShake (see http://myshake.berkeley.edu/privacy-policy/index.html).

## Acknowledgements

The Gordon and Betty Moore Foundation fund this analysis through grant GBMF5230 to UC Berkeley. The California Governor's Office of Emergency Services (Cal OES) fund this analysis through grant 6142-2018 to Berkeley Seismology Lab. We thank the previous and current MyShake team members: Roman Baumgaertner, Garner Lee, Arno Puder, Louis Schreier, Stephen Allen, Stephen Thompson, Jennifer Strauss, Kaylin Rochford, Akie Mejia, Doug Neuhauser, Stephane Zuzlewski, Asaf Inbal, Sarina Patel and Jennifer Taggart for keeping this Project running and growing. We also thank all the MyShake users who contribute to the project!